\documentclass[aps, prd, twocolumn, showkeys, groupedaddress, showpacs]{revtex4-1}
\usepackage[T1]{fontenc}
\usepackage{subfig}
\usepackage{scalerel}  
\usepackage{graphicx} 
\usepackage{bm}        
\usepackage{amssymb}   
\usepackage{lipsum}
\usepackage{lineno}
\usepackage{url}
\usepackage{amsmath,array}
\usepackage{natbib}
\usepackage{txfonts}
\usepackage[normalem]{ulem}
\usepackage{flushend}
\usepackage{hyperref}
\hypersetup{colorlinks, linkcolor={blue}, citecolor={blue}, urlcolor={blue}}
\hypersetup{
	colorlinks=true,
	linkcolor=blue,
	filecolor=magneta,
	urlcolor=blue,
}

\usepackage{tikz}
\usetikzlibrary{svg.path}
\definecolor{orcidlogocol}{HTML}{A6CE39}
\tikzset{
	orcidlogo/.pic={
		\fill[orcidlogocol] svg{M256,128c0,70.7-57.3,128-128,128C57.3,256,0,198.7,0,128C0,57.3,57.3,0,128,0C198.7,0,256,57.3,256,128z};
		\fill[white] svg{M86.3,186.2H70.9V79.1h15.4v48.4V186.2z}
		svg{M108.9,79.1h41.6c39.6,0,57,28.3,57,53.6c0,27.5-21.5,53.6-56.8,53.6h-41.8V79.1z M124.3,172.4h24.5c34.9,0,42.9-26.5,42.9-39.7c0-21.5-13.7-39.7-43.7-39.7h-23.7V172.4z}
		svg{M88.7,56.8c0,5.5-4.5,10.1-10.1,10.1c-5.6,0-10.1-4.6-10.1-10.1c0-5.6,4.5-10.1,10.1-10.1C84.2,46.7,88.7,51.3,88.7,56.8z};
	}
}
\newcommand\orcidicon[1]{\href{https://orcid.org/#1}{\mbox{\scalerel*{
				\begin{tikzpicture}[yscale=-1,transform shape]
				\pic{orcidlogo};
				\end{tikzpicture}
			}{|}}}}

\newcommand{\beq}{\begin{eqnarray}}
\newcommand{\eeq}{\end{eqnarray}}

\def\be{\begin{equation}}
\def\ee{\end{equation}}

\def\lt{\left}
\def\rt{\right}
\def \grad{\vec{\nabla}}
\def \vecB{\vec{B}}
\def\barr{\begin{array}}
	\def\earr{\end{array}}
\def\bfig{\begin{figure}}
\def\efig{\end{figure}}
\begin{document}
\title{Generating Seed magnetic field $\bf{\grave{a}}$ la Chiral Biermann battery}
\author{Arun Kumar Pandey$^{1}$ \orcidicon{https://orcid.org/0000-0002-1334-043X}\, }
\email{arunp77@gmail.com \\ arun\_pandey@prl.iitgn.ac.in}
\author{Sampurn Anand$^2$ \orcidicon{https://orcid.org/0000-0003-4346-6276}}
\email{sampurn@cutn.ac.in}
\affiliation{$^1$Department of Physics and Astrophysics, University of Delhi, Delhi 110 007, India \\
$^2$Department of Physics,	School of Basic and Applied Sciences, Central University of Tamil Nadu, Thiruvarur-610 005, India}
\begin{abstract}
	{\centering
		\bf Abstract\par}
Cosmological and astrophysical observations indicate the presence of magnetic field over all scales. In order to explain these magnetic fields, it is assumed that there exists a seed magnetic field that gets amplified by dynamos. These seed fields may have been produced during inflation, at phase transitions, or some turbulent phase of the early universe. One well-known mechanism to get the seed field is the Biermann battery, which was originally discussed in the context of generation in an astrophysical object. Requirements for this mechanism to work are (i) non-zero gradient of the electron number density and pressure, (ii) they are non-parallel to each other. In the present article, we propose a similar mechanism to generate the seed field but in inhomogeneous chiral plasma. Our mechanism works, in presence of chiral anomaly, by the virtue of inhomogeneity in the chiral chemical potential and temperature. We will discuss various scenarios where inhomogeneities in the chemical potential and temperature can arise. We found that, depending on the epoch of generation, the strength of the seed magnetic fields varies from a few nano-Gauss (nG) to a few hundred nG.
\end{abstract}
\maketitle
\section{Introduction}
\label{sec:intro}
Magnetic fields are ubiquitous in our observable Universe and are observed at all length scales, starting from  our solar system to Milky Way to galaxy clusters and superclusters and even in voids of the Large-Scale Structure (LSS). The pervading magnetic fields are expected to produce important effects on various processes including baryogenesis \cite{Giovannini:1997eg}, primordial nucleosynthesis \cite{Kernan:1995bz} and on the physics of cosmic microwave background \cite{Kunze:2014eka, Trivedi:2018ejz} (for review, see ref. \cite{Subramanian:2015lua}). Even after having so many important effects, origin of seed magnetic field remains an open ended problem in modern cosmology. It is well known that the observed magnetic fields in astrophysical structures of different sizes are produced by amplification of seed magnetic fields \cite{Furlanetto:2001gx, Bertone:2006mr}. The weakest ``seed" magnetic fields amplified by the first dynamos in the early universe could have worked at cosmological phase transition, including the Electroweak (EW) and quark confinement (QCD) phase transitions, during the Inflation or some turbulent phase in the primordial plasma due to some asymmetries \cite{Turner:1987bw, Giovannini:2000ag,  Bhatt:2015ewa, Anand:2017zpg}. Recently, the asymmetric models of the chiral plasma, where there is a finite difference in the number densities of the left-handed and right-handed massless electrons have attracted a lot of quite interesting attention \cite{Boyarsky:2012a, Anand:2018mgf, Abbaslu:2019yiy, Cassing:1999es} (for more references see the books \cite{Yagi:2005yb} and reference therein). In the present work, we have considered a mechanism to generate seed magnetic field in the early Universe due to inhomogeneities, in the chiral chemical potential and temperature, present in the chiral plasma.

The dynamics of relativistic chiral matter has been a subject of interest from both theoretical as well as experimental point of views \cite{Baym:2017whm, Li:2020dst}. The chiral matter is realized in various systems including the electroweak plasma in the early universe \cite{Kajantie:1986hq}, quark-gluon plasma in heavy ion collisions \cite{Kharzeev:2015znc}, weyl (semi-metals) \cite{Son:2012bg}, electron plasma in the neutron stars \cite{Grabowska:2015bd, Istomin:2007ge} and, in the interior neutrino medium of the core-collapse supernova explosion \cite{MEZZACAPPA1999281} (see the book \cite{Martinez:2017mg} and references therein). The chiral plasma exhibit interesting transport properties which are not seen in the normal plasma. For instance, the triangle anomaly \cite{Adler1969sx}, which arises in the context of quantum field theory, leads to chiral magnetic effect (CME) \cite{Kharzeev:2010gd} and chiral vortical effect (CVE) \cite{Kalaydzhyan:2014tk} are a few to mention. Processes related to the chiral plasma dynamics are affected by the Abelian anomaly \cite{Alekseev:1998ds, Landsteiner:2011} of the minimal standard model (MSM) and it is given by the anomaly equation \cite{joyce1997sh}: $\partial_\mu j_R^\mu=-\frac{g'^2 y_R^2}{64 \pi^2} \mathcal{F}_{\mu \nu}\bar{\mathcal{F}}^{\mu \nu}$. $\mathcal{F}$ and $\bar{\mathcal{F}}$ are the  U$_Y$(1) hypercharge field strength and their duals respectively, $j_R^{\mu}$ is the current for the right handed particles, $g'$ is the associated coupling constants and $y_R$ is the hypercharge of the right electron. The helicity of the gauge fields are shown to be related to the Chern-Simon (CS) number $N_{\rm CS}=-\frac{g'^2}{32\pi^2}\int d^3 x\epsilon_{ijk}\mathcal{F}_{ij} \mathcal{A}_k$ of the hypercharge field potential. The asymmetry in the number density of right handed electrons changes with the CS-number as $\Delta n_R=\frac{1}{2}y_R^2 \Delta N_{\rm CS}$ \cite{joyce1997sh, Cornwall:1997mj}. Origin of the asymmetry in the number density of right electrons $\delta_R= \Delta n_R/s$, where $s=\frac{2}{45}\pi^2 T^3 N_{\rm eff}$ is the entropy density, $N_{\rm eff}$ is the effective degree of freedom of the MSM, is usually attributed to the out of equilibrium processes at the GUT scale \cite{CAMPBELL1992118} (for a review, see \cite{Kolb:1990vq}). Similar expressions can also be written for the number asymmetry of the left handed particles and their antiparticles. The total number asymmetry of the chiral particles is summation of the left and right handed particle number asymmetries and it known as chiral asymmetry. Chiral asymmetry is commonly parameterized by the chiral chemical potential $\mu_5= \mu_R-\mu_L$, where‘R’ is for right handed particles and ‘L’ for the left handed particles.  The anomalous coupling of the chirality and hyper-magnetic helicity leads to an exponential growth of thermal fluctuations of these fields up to a value where it is in equipartition with the chirality. This phase of exponential growth is known as `chiral plasma instability'. The evolution of the effective magnetic helicity of the hyper-magnetic fields are given by the kinetic equations \cite{Boyarsky:2012a}: $\frac{\partial}{\partial t} \left(\mu_5+ \frac{\alpha}{\pi} \mathcal{H}_B\right) = -\Gamma_f \,  \mu_5$. In this equation, $\Gamma_f$ represents the chiral flipping rate and $\mathcal{H}_B$ is the magnetic helicity of the (hyper-charge) magnetic fields. It is to be noted here that the mentioned magnetic fields are not the standard-model electromagnetic fields, but they are the hyper-charge magnetic fields. In subsequent parts of the paper, unless we refer to this as a Standard Model magnetic field, the magnetic fields noted are hyper-charged magnetic fields. In the absence of the reactions that flip the chirality of the interacting particles, the chiral number densities are conserved. Flipping starts at temperature $T_f$, when the chiral flipping rates $\Gamma_f$ becomes equivalent to the expansion rate of the universe. For instance, for the processes at EW phase transition $T_f\sim$ TeV and for QCD scale $T_f\sim$ few GeV.  At temperatures $T> T_f$, the asymmetry in the number densities of the massless electrons remain in the thermal-equilibrium via its coupling with the hypercharge gauge bosons. Therefore, it is expected that in this regime, a helical magnetic fields are generated and they grow at the cost of chiral asymmetry in the plasma. These helical magnetic fields and the chiral symmetry of leptons support each other in the process of `inverse cascading', transferring magnetic energy from the small length scale to a large length scale. In ref \cite{Boyarsky:2012a}, it is shown that chiral asymmetry could survive till $T\sim 10$ MeV.

So far, the generation of magnetic fields are discussed mostly in a homogeneous chiral plasma~\cite{Anand:2017zpg, Rogachevskii:2017uyc}.  In the present work, we exploit chiral MHD equations used to describe the dynamics of the inhomogeneous chiral plasma to generate the seed magnetic field. We show that  inhomogeneities in the chiral chemical potential and temperature of the fluid lead to sufficiently large seed field through a Biermann battery like mechanism \cite{BIERMANN1950}. These seeds can be further amplified, by dynamos as well as instabilities in the chiral plasma, to currently observed strength of magnetic fields in voids. We call this a {\it chiral Biermann Battery mechanism}. In the present work, we have considered flat FLRW metric $\eta^{\mu\nu}$ with signature ($-, +, +, +$) and used our units in such a way that $\hbar=c=k_B=1$. This manuscript is structured as follows: section~\ref{sec:ch-bbm} provides an overview of chiral dynamics and also discusses the generation of magnetic fields by Biermann battery like mechanism. Section~\ref{sec:cond} discuss the scenario in which inhomogeneities in chiral chemical potential and temperature can arise. This also discusses the condition for the Biermann battery to be operative. Summary and the future prospects of the work is given in section~\ref{sec:result}.
\section{Chiral Biermann Battery Mechanism}
\label{sec:ch-bbm}
Generation of seed magnetic fields by a cosmic battery is commonly based on the fact that in a charge neutral universe, the positive and negative charge particles have different behavior due their mass difference. For a given pressure gradient in the gas, electrons would be accelerated much more than the ions due to it's small mass compared to ions. This leads to a current and hence an electric field $\vec{E}=-\grad p_e/(e \, n_e)$. If the curl of the thermally generated electric field has non zero value, then from Faraday's law of induction, magnetic field can grow. The resulting battery is termed as Biermann battery mechanism~\cite{BIERMANN1950}. This mechanism is mostly explored in the context of stellar objects and early universe processes at the time of recombination. However, before the recombination epoch, a similar mechanism can be operative in chiral plasma provided there is an inhomogeneity in chiral chemical potential and temperature. Before proceeding further, we provide a brief overview of relativistic MHD equations required to describe the dynamic of chiral fluids. Later, we use them to derive an equation which looks similar to Biermann battery.
\subsection{Overview of Chiral fluid}
\label{sec:ch-review}
To provide an overall description of chiral plasma, we assume that an external magnetic field is present in the beginning. Later on, we will come to the case where initial magnetic field is absent and seed field is generated. In presence of an external magnetic field the hydrodynamic equations that govern the time evolution of the anomalous chiral fluid are given by the following set equations \cite{Yamamoto:2015ria}
\begin{eqnarray}
\nabla_\mu\, T^{\mu\nu} & = & F^{\nu\lambda}\, j_\lambda\,  ,
\label{eq:StressCons} \\
\nabla_\mu\, j^\mu  & = &  0\, ,
\label{eq:VCons} \\
\nabla_\mu\, j_5^\mu  &= &  C\, E_\mu\,B^\mu
\label{eq:ACons}
\end{eqnarray}
where $T^{\mu\nu}$ is the energy momentum tensor of an ideal fluid, $F^{\nu\lambda}$ is the electromagnetic field strength tensor. The electric and magnetic field four vectors are represented by $E^\mu$ and $B^\mu$ respectively. The vector current is given by $j^\mu = j^\mu_R +j^\mu_L$ and the chiral current is represented by $j_5^\mu = j^\mu_R -j^\mu_L$. The chiral anomaly coefficient is denoted by $C$. 

In the state of local equilibrium, the energy momentum tensor $T^{\mu\nu}$, the vector current $j^\mu$ and the chiral current $j^\mu_5$ can be expressed in terms of the four velocity of the fluid $u^\mu$, energy density $\rho$, vector charge density $n_{\rm v}$ and axial charge density $n_5$. In absence of electromagnetic fields, the local equilibrium reached at a length scale of spatial variation of chemical potential $\mu$, i.e. $\ell_{\rm LTE}\ll \ell_\mu \sim  \mu (\vec{x})/ |\vec{\nabla} \mu(\vec{x})|$ (here $\ell_\mu$ is the scale over which chemical potential varies significantly). Thus, local distribution function of the fermions is given by the local expression $f_i^{\rm eq}(t, \vec{x}, \vec{p}) = \left[\text{Exp} \left(\frac{\epsilon_p-\mu_i(t, \vec{x})}{T(t, \vec{x})} \right) + 1 \right]^{-1}$, where $\epsilon_p=c|\vec{p}|$ (here `$i$'=right/left handed particles). Spatial variation of electromagnetic fields and matters occurs at much larger than $\ell_{\rm LTE}$. Since the chiral anomaly relation is local, the electric and chiral chemical potentials should be space-time dependent. The relation between axial charge density $\rho_5$ and the zeroth component of axial current $j_5^\mu$ is given by $\rho_5(t, \vec{x})=\langle j_5^0(t, \vec{x})\rangle_{T, \mu_5}$ which is valid for $\mu_5\ll T$ \cite{Boyarsky:2015faa}. For small deviation from local equilibrium, vector current and chiral current respectively takes the form \cite{Son:2009tf, Gorbar2016}
\begin{eqnarray}
   j^\mu & = & n_{\rm v}\,u^\mu -  \frac{\sigma}{2}T\, \Delta^{\mu\nu}
  \partial_\nu \left(\frac{\mu_{\rm v}}{T}\right) + \sigma E^\mu + \xi_{\rm v}\omega^\mu +
  \xi_{\rm v}^{(B)}B^\mu \, ,   \label{eq:VectorCurrent} \\
    j_5^\mu & = & n_5\,u^\mu -  \frac{\sigma}{2}T\,
  \Delta^{\mu\nu} \partial_\nu \left(\frac{\mu_{5}}{T}\right)
  + \xi_{\rm 5}\omega^\mu + \xi_{5}^{(B)}B^\mu\, ,
  \label{eq:AxialCurrent}
\end{eqnarray}
here $\Delta^{\mu\nu}=(\eta^{\mu\nu}+u^\mu u^\nu)$ is the projection operator, $n_{\rm v,5} = n_{R} \pm n_L$, $\mu_{\rm v,5} = \mu_R \pm \mu_L$, $\xi_{\rm v,5} = \xi_R \pm \xi_L$, $\xi^{(B)}_{\rm v,5} = \xi^{(B)}_R \pm \xi^{(B)}_L$, and $\sigma$ is the conductivity. The vorticity four vector is represented by $\omega^\mu$. The mathematical expression of the transport coefficients $\xi_{{\rm v}, 5}$ and $\xi_{{\rm v}, 5}^{(B)}$ in equations (\ref{eq:VectorCurrent}) and (\ref{eq:AxialCurrent}) are calculated by many authors and it is shown that these terms are not only allowed but they are required for anomalies \cite{Son:2009tf, Landsteiner:2011, Bhatt:2015ewa}. The second term in equations (\ref{eq:VectorCurrent}) and (\ref{eq:AxialCurrent}) arise only when there is inhomogeneity in either chemical potential or temperature or in both. For the present study, inhomogeneity in both chemical potential as well as temperature are important. The total current ($j^\mu_{\rm tot} = j^\mu + j_5^\mu$) from the right handed chiral particles is given by
\begin{equation}
	j^\mu_{\rm tot} = 2\, n_{\rm R}\,u^\mu + \sigma E^\mu
	+ 2\,\xi_{\rm R}\, \omega^\mu + 2\,\xi_{\rm R}^{(B)}\, B^\mu
	- \sigma\,T\,\Delta^{\mu\nu}\partial_\nu\left(\frac{\mu_{\rm R}}{T}\right).
	\label{eq:jtotR}
\end{equation}
The  coefficients $\xi_{\rm R}$ and $\xi_{\rm R}^{(B)}$ are given as \cite{Neiman2011, Son:2009tf}
\begin{eqnarray}
	\xi_{\rm R} & = & C\,\mu_{\rm R}^2\,\lt[1-\frac{2\,n_{\rm R}\,\mu_{\rm R}}{3\,(\rho + p)}\rt]\, +
	\frac{D\,T^2}{2}\lt[1-\frac{2\,n_{\rm R}\,\mu_{\rm R}}{(\rho + p)}\rt] \, ,
	\label{eq:xi} \\
	\xi_{\rm R}^{(B)} & = & C\,\mu_{\rm R}\,\lt[1-\frac{n_{\rm R}\,\mu_{\rm R}}{2\,(\rho + p)}\rt]\, -
	\frac{D}{2}\lt[\frac{n_{\rm R}\,T^2}{(\rho + p)}\rt]\, .
	\label{eq:xiB}
\end{eqnarray}
The second term in these equations are uniquely fixed by the requirement on the entropy current  $s^\mu$ to satisfy $\partial_\mu s^\mu \geq 0$ \cite{Son:2009tf}. The coefficient `$D$' in above equations cannot be derived solely from hydrodynamics \cite{Neiman2011}, and it is a manifestation of additional microscopic properties of the chiral degrees of freedom. Values of the coefficient $D$ is derived by considering gauge-gravitational duality by many authors \cite{Landsteiner:2011}. In the simplest case of non-interacting chiral fermion, values of the coefficients $C$ and $D$ are given by $C=1/4\pi^2$ and $D=1/12$.
\subsection{\label{sec:ch-BBM} Seed magnetic field generation}
\noindent
Using the expression for $\vec j_{\rm tot}$, given in equation (\ref{eq:jtotR}) and the  Maxwell's equation $\grad \times \vecB = \vec j_{\rm tot}$, we get
\begin{eqnarray}
\label{eq:dcrossB}
\grad \times \vec{B} & = &  2 n_{\rm R}\vec{v} - \sigma T \left[\vec{\nabla} \left(\mu_{\rm R}/T\right) + \vec{v} \partial_t \left(\mu_{\rm R}/T\right)  + \vec{v} (\vec{v}\cdot\vec{\nabla}) \lt(\mu_{\rm R}/T\rt)\right]\nonumber \\
& + & \sigma(\vec{E} +\vec{v}\times \vec{B})  +
2\,\xi_{\rm R}\vec{\omega} + 2\,\xi_{\rm R}^{(B)}\vec{B}.
\end{eqnarray}
To obtain the evolution equation for the magnetic fields, we first eliminate $\vec{E}$ from the above equation as
\begin{eqnarray}
\label{eq:dcrossB-2}
\vec{E} & = & -\vec{v}\times \vec{B}+\frac{\grad \times \vec{B}}{\sigma}-\frac{2 n_{\rm R}}{\sigma}\vec{v} -\frac{2\,\xi_{\rm R}}{\sigma}\vec{\omega} - \frac{2\,\xi_{\rm R}^{(B)}}{\sigma}\vec{B} \nonumber \\
& + & T \left[\vec{\nabla} \left(\mu_{\rm R}/T\right) + \vec{v} \partial_t \left(\mu_{\rm R}/T\right)  + \vec{v} (\vec{v}\cdot\vec{\nabla}) \lt(\mu_{\rm R}/T\rt)\right].
\end{eqnarray}
Now taking curl of the above equation and using $\vec\nabla \times\vec E = -\frac{\partial \vec{B}}{\partial t}$, the evolution equation for $\vec{B}$ field is given as:
\begin{widetext}
\begin{eqnarray}
\frac{\partial \vecB}{\partial t}& = & 
\underbrace{\frac{1}{\sigma}\,\nabla^2\vec{B}}_{\rm I} 
+ \underbrace{\grad\times(\vec{v}\times \vec{B})}_{\rm II} 
+ \underbrace{\frac{2}{\sigma}\,\left[n_{\rm R}\,\vec{\omega} + \vec{\nabla} n_{\rm R} \times \vec{v}\right]}_{\rm III}
+ \underbrace{\frac{2}{\sigma}\left[\vec{\nabla} \xi_{\rm R}\times \vec{\omega} + \xi_{\rm R} \vec{\nabla}\times \vec{\omega}\right]}_{\rm IV}  
+ \underbrace{\frac{2}{\sigma}\left[\vec{\nabla} \xi_{\rm R}^{(B)} \times \vec{B} + \xi_{\rm R}^{(B)} (\vec{\nabla}\times\vec{B})\right]}_{\rm V} 
+ \underbrace{\frac{\vec \nabla\sigma}{\sigma^2}\,\times\left(\vec\nabla\times\vec{B}\right)}_{\rm VI} 
\nonumber \\ 
& &
-\underbrace{\frac{2n_{\rm R}}{\sigma^2}\left(\vec{\nabla}\sigma \times \vec v\right)}_{\rm VII}
-\underbrace{\frac{2\xi_{\rm R}}{\sigma^2}\left(\vec{\nabla}\sigma \times \vec \omega\right)}_{\rm VIII}
-\underbrace{\frac{2\xi^{\rm (B)}_{\rm R}}{\sigma^2}\left(\vec{\nabla}\sigma \times \vec B\right)}_{\rm IX}
-\underbrace{\frac{1}{T}\left[\vec{\nabla} T \times \vec{\nabla}\mu_{\rm R}\right]}_{\rm X} 
- \underbrace{\vec{\nabla} \times \left[T\, \left\{\vec{v}\,\partial_t\,\left(\mu_{\rm R}/T\right)  + \vec{v}\, (\vec{v}\,\cdot \vec{\nabla}) \left(\mu_{\rm R}/T\right)\right\}\right]}_{\rm XI}.
\label{eq:magevol}
\end{eqnarray}
\end{widetext}
%
%
%
It is important to note here that this equation is valid for the case when chiral plasma is inhomogeneous. In the case of homogeneous chiral plasma, all terms with $\vec\nabla \sigma$, $\vec\nabla n_R$, $\vec\nabla \xi_R$, $\vec\nabla \xi_R^{(B)}$, $\vec\nabla T$ and $\vec\nabla \mu_R$ will vanish. Before proceeding further, we estimate the order of magnitude of each term in the right hand side of the above equation. In order to do so we take $\sigma\sim T/4\pi\alpha$ \cite{Baym:1997gq}, $n_{\rm R} \sim \mu T^2/6$, 
$\xi_{\rm R}\sim \lt(\mu^2/\pi\rt)\sqrt{\alpha/4\pi}$ and $\xi_{\rm R}^{(B)}\sim 2\alpha \mu/\pi$. If $L$ is the length scale of interest, term by term order in equation (\ref{eq:magevol}) is: ${\rm I}  \sim 4\pi\alpha\lt(\frac{B}{T^2}\rt)\lt(\frac{T}{L^2}\rt) = {\rm VI}$, ${\rm II} \sim \frac{B\,v}{L}= \lt(\frac{B}{T^2}\rt)\lt(\frac{T^2}{L}\rt) v$,
${\rm III} \sim \frac{2\,n_{R}\,v}{\sigma L} \sim \frac{4\pi}{3}\alpha\lt(\frac{\mu}{T}\rt)\lt(\frac{T^2}{L}\rt)\,v = {\rm VII}$,
${\rm IV} \sim \frac{2\xi_{\rm R}\,v}{\sigma L^2}\sim \frac{4}{\pi}\alpha^{3/2}\lt(\frac{\mu}{T}\rt)^2\lt(\frac{T}{L^2}\rt)\, v = {\rm VIII}$,
${\rm V} \sim \frac{2\,\xi_{\rm R}^{(B)} B}{\sigma L} \sim 16\alpha^2\lt(\frac{\mu}{T}\rt)\lt(\frac{B}{T^2}\rt)\lt(\frac{T^2}{L}\rt) ={\rm IX}$,
${\rm X} \sim\lt(\frac{\mu}{T}\rt)\lt(\frac{T}{L^2}\rt) ={\rm VIII}$, 
${\rm XI} \sim \lt(\frac{\mu}{T}\rt)\lt(\frac{T}{L^2}\rt)\,v^2\,$.  Since $\alpha \sim 10^{-2}$, $\frac{\mu}{T}\sim (10^{-4}-10^{-6})$, $v\ll 1$ we can ignore all other terms compared to I, II and X. Further at temperature scale of our interest, variation in conductivity is also small and VI, VII, VIII and IX term can be dropped. Hence, equation (\ref{eq:magevol}) will reduced to following form with above approximation
\begin{equation}
\frac{\partial \vecB}{\partial t} = \frac{1}{\sigma}\,\nabla^2\vecB + \grad\times(\vec{v}\times \vecB)-\frac{1}{T}\lt[\vec \nabla T \times \vec \nabla \mu_{\rm R}\rt].
\label{eq:diffusion}
\end{equation}
Above equation represents the magnetic induction equation for the magnetic fields in the case of inhomogeneous chiral plasma. The first two terms on the right hand side represents the diffusion and the convection respectively. First term signifies the transport of the magnetic field via diffusion. However, second term describes the magnetic field in a conducting fluid changes with time under the influence of a velocity field $v$. In absence of initial electromagnetic fields {\it i.e.} $\vec{E} =0 =\vec{B}$, equation (\ref{eq:diffusion}) reduces to
\begin{equation}
\frac{\partial \vecB}{\partial t} = 
-\frac{1}{T}\lt[\vec \nabla T \times \vec \nabla \mu_{\rm R}\rt] .
\label{eq:ch-bbm1}
\end{equation}
It is important to note here that seed magnetic fields are produced via this mechanism only in the case of inhomogeneous chiral plasma. Along with this, following conditions should also be satisfied: $\vec \nabla T \times \vec \nabla \mu_{\rm R} \neq 0$ or a non-parallel components of $\vec \nabla T$ and $\vec \nabla \mu_{\rm R}$. This equation looks exactly like Biermann mechanism. Therefore, we call this mechanism as a {\it Chiral Biermann battery} mechanism. 
\section{Conditions for the Chiral battery}
\label{sec:cond}
For Chiral Biermann mechanism to work, following conditions must be satisfied i). $\vec \nabla \mu_5\neq0$,  $\vec \nabla T\neq 0$ and ii). $\vec \nabla T \nparallel \vec \nabla \mu_{\rm R}$. Here we have discussed three important scenarios where all the three conditions are satisfied. 
\begin{itemize}
    \item One of the most promising scenario to achieve all three conditions is the first order phase transition or during any turbulent phase of early universe \cite{Baym:1996lm}. Our universe has gone through several phase transitions (PT) including electroweak, at around $T_{\rm EW} \sim 100$ GeV, and the QCD phase transition occurring around the critical temperature $T_{\rm QCD} = 150$ MeV. Although the opinion is divided, various arguments raise the possibility that these transitions might be first order. In this work, we will assume that the QCD phase transition is first order. If the QCD transition is first order then the Universe has to cool somewhat below critical temperature $T_{\rm QCD}$ before any regions of hadron matter appear. The universe supercools a finite amount before the appearance of small nucleation sites. These are bubbles of hadronic phase which consist mostly pions. It is important to highlight the fact that there are two different time scales involved, namely (i) the QCD time scale which is of the order of $\tau _{\rm QCD} \sim 1/T_{\rm QCD}$ and the Hubble time scale which is $\tau_{\rm H}\sim 10^{19}/T_{\rm c}\gg \tau_{\rm QCD}$. Thus, the nucleation is a local phenomena. After nucleation, bubble grows explosively like a deflagration bubble. For small supercoolings (of the order $2\% $ {\it i.e.} $T_s\sim 0.98~T_{\rm QCD}$) the deflagration front travels slowly {\it i.e.} $v_{\rm front} \ll c_{s} =1/\sqrt{3}$ \cite{Kajantie:1986hq}. However, the front is preceded by a supersonic shock which moves with a velocity $v_{\rm sh} > c_{\rm s}$. The propagation of shock leads to heating and compression in the quark matter. With increasing time more and more bubbles are nucleated, they grow and the shock fronts preceding the bubbles begin to collide. At this stage, universe enters into a turbulent phase. If the supercooling is small then the turbulence dies out, the Universe outside the hadron bubbles is reheated to $T_{\rm QCD}$ and the explosive bubble growth is halted. At this time hadronic bubbles are roughly 1/10 of the average distance between nucleation sites. In the case of small supercooling, collisions between two shock fronts and between a shock and a deflagration front may lead to inhomogeneity in the temperature as well as chemical potential. These inhomogeneities exist over a scale of coexisting phases. Moreover, it was assumed that the hardonic phase are spheres of the same size. This is only an approximation. In fact, there is a complex distribution of sizes due to the fact that the nucléation sites do not appear at exactly the same time. Also, their shapes are not exactly spherical and may include ripples when surface tension becomes unimportant. Thus, when the shock fronts collide, a turbulent phase begins and vorticity is generated \cite{Quashnock:1989aj}. During this phase, all three conditions required for the generation of seed field are met. An estimate of the generated seed field can be given as follows: The duration of QCD phase transition $t_{\rm pt} \sim 0.22\tau_{\rm H}\simeq 43\mu$s, temperature will be inhomogeneous over the scale of coexisting phase and so is the chemical potential. $\Delta T /T_{\rm QCD} \sim (\ell /\tau_{\rm H})^2$, where $\ell$ is the scale over which the temperature and chemical potential will be inhomegeneous \cite{Kajantie:1986hq}. This scale is typically the size of different coexisting phases. Using these number, we can estimate the strength of seed field generated at the source as follows: $B_{\rm QCD} \sim t_{\rm pt}\times \left(\frac{\Delta T}{T_{\rm QCD}}\right) \left(\frac{\Delta \mu_5}{T_{\rm QCD}}\right)\left(\frac{T_{\rm QCD}}{\ell^2 }\right)\sim 0.26$ nG.
     \item Another interesting scenario which can generate inhomogeneity in temperature and chemical potential is that of the Inhomogeneous QCD phase transition proposed in ref.~\cite{Ignatius:2000cz}. This is possible when temperature is inhomogeneous. The inhomogeneous temperature depends on two parameters, (i) density perturbation $\Delta_T^{\rm (rms)}$ and (ii) the temperature interval of nucleation $\Delta_{\rm nuc}$. It has been proposed that, when $\Delta_T^{\rm rms}> \Delta_{\rm nuc}$, the nucleation of the bubbles at a given time will be inhomogeneous \cite{Ignatius:2000cz}. Since, the inflation produced density perturbation which leads to temperature fluctuation of the order of $\Delta_T^{(\rm rms)} \sim 10^{-5}$ and results of the lattice simulation with quenched QCD (no dynamical quarks) give the value of (dimensionless) temperature interval of nucleation $\Delta_{\rm nuc} \sim 10^{-6}$, the nucleation is thus inhomogeneous. Initially, cold spheres of diameter $\ell_{\rm smooth} \sim 10^{-4} d_{\rm H}$ (where $d_{\rm H}=c/H\sim 10$ km is the Hubble distance at the QCD transition \cite{suh:1998smi}) with equal and uniform temperature are distributed randomly which is $\Delta_T^{(\rm rms)}~\times~T_{\rm QCD}$ less than the rest of the uniform universe. When the temperature of the cold spot decreases to the value of actual nucleation temperature $T_{\rm n}$, homogeneous nucleation takes place within it. However, the Hubble expansion would result in the cooling of the universe and would take $\Delta t_{\rm cool} =(\Delta_T^{(\rm rms)}/3c_s^2)~\tau_H$ time to cool down to $T_{\rm n}$. Inside each cold spot, there will be large number of tiny hadron bubbles. These bubbles merge within $\Delta t_{\rm cool}$ if $\Delta_{\rm nuc} < (v_{\rm def}/v_{\rm heat}) \Delta_T^{\rm rms}$, where $v_{\rm def}$ is the speed of deflagration front and $v_{\rm heat}$ is the effective speed by which released latent heat propagates to  stop nucleation. The length scale of temperature propagation is determined by the latent heat released in the cold spot which propagates in all directions and is given as $\ell_{\rm heat} = 2 v_{\rm heat} \Delta t_{\rm cool}$ (which is of the order of few meters \cite{Suh:1998isgj}). Thus, in this scenario we can have temperature gradient of the order of $\Delta T_{\rm QCD}/\ell_{\rm heat}\sim 10^{-1}-10^{-4}$ MeV/km (when $\ell_{\rm heat}/\ell_{\rm smooth}=1, 2, 5, 10$). In this scenario, again obtained values of the magnetic fields strength is in the range of $\sim$ nG.
     \item There is one more viable scenario, around electroweak scale, where all the three condition can be satisfied is when few hyper-magnetic modes grow exponentially in a chiral plasma. It has been shown that in chiral plasma, due to finite chiral asymmetry, the quantum effects leads to the production of hyper-charge magnetic fields. Few modes of these fields show an exponential growth due to parity odd interaction of the fermions to the Abelian fields. The exponential growth of those modes with wave number $k$ during a chiral plasma instability  occurs at a time scale of $t\sim\sigma/k(4\mu-k)$, where mode with wave number $k_{\rm max}=2\mu$ have maximum growth rate~\cite{Akamatsu2013}. During this phase, chiral plasma goes through a turbulent evolution. The time scale of the maximum growth rate corresponding to chemical potential $\mu_{\rm max}=100 T/ 4\mu^2$. The requirement that the growth time scale should be smaller than the Hubble time $\tau_H$ gives constraint on $\mu/T=\delta$ as $\delta =2\times 10^{-6}\,  (T/T_f)^{1/2}$. If the temperature at which left-right asymmetry is generated at a temperature $T< T_f$, time available for the generation of the magnetic fields is $\Gamma_f^{-1}$, rather than Hubble time. This means, we can consider asymmetry $\delta\geq 2\times 10^{-6}$ \cite{joyce1997sh, Neronov:2020nyt}. In this case, for a typical values of $T\sim 100$ GeV, $\ell_\mu\sim \ell_{\rm heat} \sim$ GeV$^{-1}$ and $t_{\rm EW}\sim $ GeV$^{-1}$ and $\delta_{\rm EW}\sim 10^{-6}$, strength can be calculated by using  $B(t_{\rm EW}) \sim \delta_{\rm EW} \times (T/\ell_{\rm heat}) (t/\ell_\mu)\sim 10^{-3}$ nG. 
\end{itemize}
\section{\label{sec:result} Summary and future prospect of the work}
In this work, we have used Biermann battery like mechanism to explore the possibility of generation of seed magnetic field in the early universe. Our mechanism works when chiral chemical potential and the temperature have spatial dependence. Magnetic fields generation via this mechanism may work at phase transitions or during the turbulent phase of exponential growth of the chiral modes (this is known as chiral plasma instability) in presence of finite amount of the chiral asymmetry. The strength of the generated magnetic fields via this mechanism are of the order of $\sim$nG. The helical magnetic fields produced by this mechanism subsequently evolves through various turbulent phases which preserves the helicity and ultimately produce the primordial standard model magnetic fields surviving till present epoch. Once the primordial helical magnetic fields are generated by chiral Biermann battery mechanism at length scale larger than the ``frozen-in'' scale $L_f\sim \sqrt{t/4\pi\sigma}$ (here $t$ is the cosmic time), the first and second term in equation \eqref{eq:diffusion} dominates. At high temperature conductivity of the chiral plasma, ($\sigma \sim T/\alpha \ln(1/\alpha)$, see the ref. \cite{Baym:1997gq}), is  very large  and hence dissipation by first term in equation \eqref{eq:diffusion} can be ignored. The dynamics of the fields are solely governed by the convection term. At length scales $\ell> L_f$, magnetic fields are coherent and are said to be frozen in. It means that, helicity of the magnetic fields are conserved both locally and globally. When the fluid velocity is small, it is also possible that the first term in equation \eqref{eq:diffusion} dominates over the second term . In this case magnetic fields dies out exponentially and helicity is no longer conserved. If fluid velocity is not small, the evolution of the fields is governed by both terms. In this situation, it has been argued that the local helicity of the magnetic fields changes due to the reconnection of field lines and therefore, they are not conserved. However, the global helicity, a summation of random local changes, remain conserved \cite{Taylor:1974jb}. Reynolds number $R_M=4\pi\sigma l v\gg1$ decides the conservation of global helicity at a characteristic length scale $l\sim 1/e^2T$ (which characterizes the gauge field configurations such as the sphalerons). Whether this criteria is satisfied, depends on the dynamics of the fluid during baryogenesis and requires significant departure from the thermal equilibrium. Under these circumstances, magnetic fields evolve conserving the global magnetic helicity even though the field is not frozen in.

So far in this work, we have focused our attention on the production of seed field in a inhomogeneous chiral plasma. To glean a complete picture of the generated magnetic field and processes involved in it, we need to study the evolution of generated magnetic field to today's epoch and the power spectrum of the magnetic field. These would be the topics of future studies.
\section*{Acknowledgment} 
Authors thank K. Subramanian and T. R. Seshadri for their valuable comments and suggestions on the manuscript. AKP is financially supported by Dr. D.S. Kothari Post-Doctoral Fellowship, under the Grant No. DSKPDF Ref. No. F.4-2/2006 (BSR)/PH /18-19/0070. AKP also wishes to thanks facilities provided at ICARD, Department of Physics and Astrophysics, University of Delhi, India. 
\bibliographystyle{apsrev4-1}
\bibliography{bbm-ref} 
\end{document}